# Investigation of (CsMA)NaBiX$_6$ (MA= methylammonium; X=Cl, Br, I) Organic-Inorganic Hybrid Double Perovskites for Optoelectronic Applications: A First Principles Study

**A. Johnson** [a*], **B.A. Ikyo** [a], **T. J. Ikyumbur** [a], **F. Gbaorun** [a] **and A. A. Nathan** [b, c].

[a]Department of Physics, Benue State University, Makurdi, Nigeria.
[b]Department of Material Science, Centro de Investigacion ´en Materiales Avanzados, S.C., Miguel de Cervantes 120, Complejo Industrial Chihuahua, 31109, Chihuahua, CHIH, Mexico.
[c]Department of Physics, Joseph Sarwuan Tarka University, Makurdi, Nigeria.

**Corresponding Author's Email:* ajohnson17@bsum.edu.ng

**Abstract**
Organic-inorganic hybrid double perovskites have attracted increasing interest in the commercialization of lead-free and nontoxic perovskites due to their unique optoelectronic properties compared with inorganic perovskites. In this study, the structural, electronic, optical and properties of 3 Pb-free compounds (CsMA)NaBiX$_6$ (MA= methylammonium; X= Cl, Br, I) were simulated using first-principles density functional theory (DFT). Results show that the investigated compounds are semiconductors with tunable bandgaps characteristics that can be used in devices like light emitting diodes, and predict the suitability of the (CsMA)NaBiI$_6$ organic-inorganic hybrid double perovskite in this study for optoelectronic applications owing to its high absorption coefficient (in the order of $10^6 cm^{-1}$), dielectric constant (approx.3.24), and refractive index (2.50) as well as its stability as revealed by its high formation energy. Additionally, the high absorption coefficient, high refractive indices and dielectric constants of the investigated materials posits that they have a number of optoelectronic applications including photovoltaic cells among others.
**Keywords:** Lead free perovskites, Hybrid semiconductors, Optoelectronics, Thermalproperties

## 1 Introduction

Increasing interest in recent time has been drawn to novel materials such as double perovskites with the basic formula A$_2$B′B″X$_6$, where A is a cation typically, Cs$^+$ or methylammonium (MA$^+$), B′ is a trivalent metal cation such as Bi$^{3+}$ or Sb$^{3+}$, B″ is a monovalent metal usually Cu$^+$, Ag$^+$, or Au$^+$ and X isusually a halide such (Kim & Kim, 2018; Sani & Shafie, 2019; Volonakis et al., 2016; Zhang, Liao, & Yang, 2019) . This is largely attributed to the fact that these materials offer suitable replacements for lead as well as the solving the instability defects of lead-halide perovskites(Volonakis et al., 2016). Nonetheless, better optoelectronic properties have been observed forMA based organic–inorganic hybrid perovskites than Cs contained inorganic perovskites in addition to depicting similar optoelectronic properties with Cs$_2$BiAgI$_6$, Cs$_2$SbAgI$_6$ and MAPbI$_3$ (Roknuzzaman et al., 2019; Volonakis et al., 2016).

Among all reviewed hybrid double perovskites (Filip, Filip, & Volonakis, 2018; Meng et al., 2017), the use of a noble metal is evident in additionto employing Cu or Ag as B″ site element, however, due to depletion of the copper ore (Rötzer & Schmidt, 2018) and the high cost of silver, in this study, Na as a B″ site element has been used while investigating different combination to find suitable replacement for Pb in double perovskites for different applications. The choice of Na is attributed to the fact that within the coordinate number 6, Ag$^+$ with 1.15 Å ionic radius can be substituted perfectly with Na$^+$ (1.02 Å) because of the similarity of their ionic radius (Shuai & Ma, 2016; Volonakis et al., 2016). To maximize availability of these novel materials as well as curb cost associated with noble metal-based precursors, the investigation of (CsMA)NaBiX$_6$ (MA= methylammonium; X=Cl, Br, I) organic-inorganic hybrid double perovskites for optoelectronic applications: a firstprinciples study is presented.

## 2 Materials and Method

### 2.1 Computational Details
The plan waves-pseudopotentials (PW-PPs) method as implemented in the Quantum ESPRESSO (QE) (Lab, 2021) code based on DFT was used to calculate the ground states of the perovskites in the present study. The ultrasoft pseudopotential including core correction was used for all calculations. The local density approximation (LDA) of Perdew-Berke-Ernzerhof (PBE) (Ropo, Kokko, & Vitos, 2008) was used to calculate the optimized structures at 4×4×4 $k$ points grid. In order to optimize the crystal structure and estimate the ground state energy, the Broyden-Fletcher-Gordfarb-Shanno (BFGS) minimization technique was employed. To achieve energy eigenvalue convergence, the wavefunctions were expanded in plane waves with a cutoff energy of 30Ry and the charge density set at 180 Ry. The Monkhorst-Pack scheme $k$ point of 4×4×4 was employed in sampling the Brillouin zone for obtaining the best convergence and total energy value for all calculations associated with LDA-PBE in this study. The cell stresses were at least $10^{-6} Ry/bohr^3$ and the residual forces on each atomic site were at least $10^{-4} Ry/bohr$.

### 2.2 Formation Energy Calculation
For synthesis of the investigated compounds, the thermal stability of these perovskites is calculated. The formation energies, $E_{form}$ for the perovskites were evaluated to indicate the thermal stability of these compounds and the values are presented in Table 2. Moreover, the energy of formation is calculated using Equation (1):

$$E_{form} = E_{tot} - \Sigma_x E_{tot}(x) \qquad (1)$$

Where, $E_{form}$ is the formation energy of the perovskite; $E_{tot}$ is the total energy of the perovskite compound and $\Sigma_x E_{tot}(x)$ is the sum of the total energy components of the perovskite compound which were obtained from the energy calculation in the optimization of the compounds. As can be inferred from Table 2, the negative values of formation energy depict that the perovskites are energetically stable and be suitably synthesized experimentally.

## 3 Results and Discussion
### 3.1 Structural properties
The optimized lattice parameters for the simulated hybrid double perovskites are listed in Table 1. These Bi-based hybrid inorganic and organic double perovskites show increasing trend in order from X= Cl, Br and I.

Table 1. Calculated lattice constants and bandgaps for hybrid double halide perovskites (CsMA)NaBiX$_6$ (X = Cl, Br, I).

| Compounds | a (Å) | Bandgap (eV) (LDA-PBE) |
|---|---|---|
| (CsMA)NaBiCl$_6$ | 7.82 | 3.69 |
| (CsMA)NaBiBr$_6$ | 8.24 | 3.07 |
| (CsMA)NaBiI$_6$ | 8.85 | 2.36 |

Results in Table 1 showed that the bandgap energy increases with reducing lattice constants in accordance with the observation by (Greenman, Williams, & Kioupakis, 2019; Roknuzzaman et al., 2017) that the replacement of halogen atom (X) by a lighter and smaller halogen atom reduces the lattice parameter. In essence, the halogens increase in atomic size down the group in Periodic Table (Kragh, 1963; Yao et al., 2021) and as such, increasing the halide concentration resulting in decrease of the bandgap energy (Johnson, A, Gbaorun, F and Ikyo, 2022). The band gap of the investigated perovskites can be tuned by changing the halogen contents which is a characteristic of light emitting diode (LEDs) (Roknuzzaman et al., 2017, 2019).

### 3.2 Electronic properties
Figures 1-3 depict the electronic band structures for the hybrid double perovskite compounds. These hybrid double perovskites were predicted to be of indirect band gap as the conduction band minima (CBMs) and valence band maxima (VBMs) appear at different points on high symmetric path of the BZ similar to their inorganic double perovskites, $Cs_2NaBX_6$(B = Bi, Sb; X= Cl, Br, I) (Shuai & Ma, 2016). The CBMs were shown to appear at the X point of the high symmetric path of the BZ. On the other hand, the VBMs of these materials were found to be at the M point with exception of (CsMA)NaBiI$_6$ with VBM at Γ. The calculated band gap values are presented in Table 1 from which it can be inferred that the iodide hybrid double perovskite ((CsMa)NaBiI$_6$ (2.36 eV)) exhibit the smallest band gap value. These values are comparable with 2.19 eV of $Cs_2AgBiBr_6$ obtained through measurement techniques by (McClure, Ball, Windl, & Woodward, 2016). All the perovskites in the present method gave band gap values of 2.36 eV, 3.69 eV and 3.07 eV for (CsMa)NaBiI$_6$, (CsMA)NaBiCl$_6$ and (CsMA)NaBiBr$_6$ respectively. These agree with Volonakis *et al*., in 2016 for $(MA)_2BiAgCl_6$ (2.7 eV) and MAPbCl$_3$ (3.0eV) by McClure *et al*., 2016. However, (CsMA)NaBiCl$_6$ perovskite with 3.7 eV in the present study depict a slight deviation which could be as a result of its small value of lattice parameter as observed by Greenman *et al*., (2019).

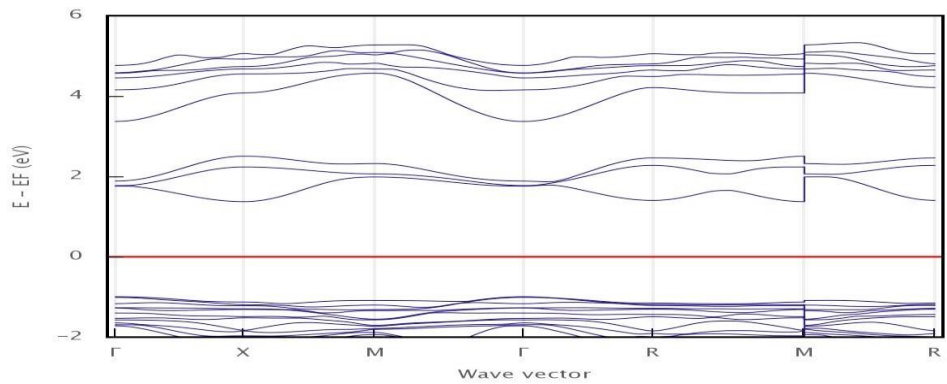

Figure 1. Electronic band structure of (CsMA)NaBiI$_6$. Bandgap energy
Eg = 2.36 eV with CBM at X and VBM at Γ in the LDA-PBE approximation.

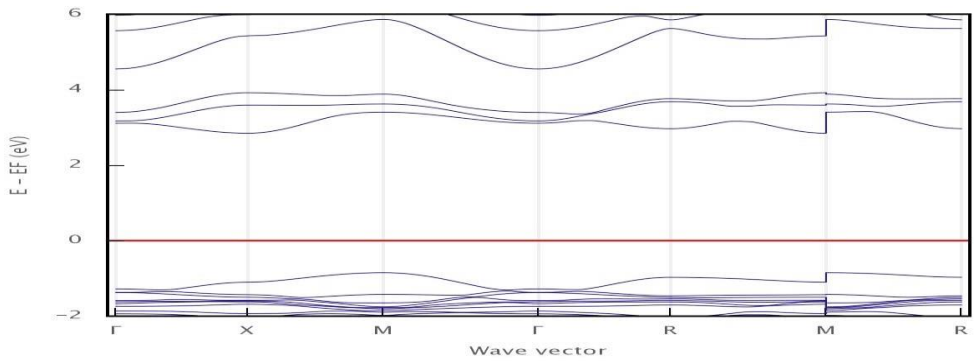

Figure 2. Electronic band structure of (CsMA)NaBiCl$_6$. Bandgap energy
Eg = 3.69 eV with CBM at X and VBM at M in the LDA-PBE approximation.

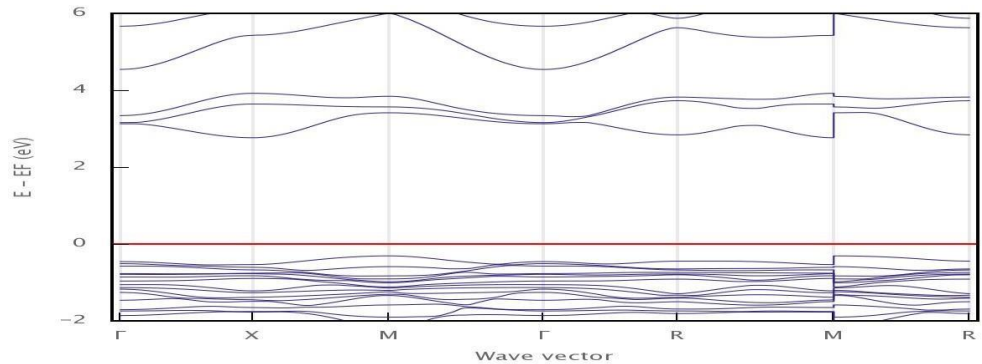

Figure 3. Electronic band structure of (CsMA)NaBiBr$_6$. Bandgap energy
$E_g$ = 3.07 eV with CBM at X and VBM at M in the LDA-PBE approximation.

Result in Figure 4 of the total and projected density of states (DOS) of the (CsMA)NaBiI$_6$ perovskite shows that the contribution to the total DOS towards VBM is by the Sb-5p states and I-5p states. The contribution to the total DOS towards CBM is by the I-5p states while the Na-3s states gave a partial contribution to the total DOS towards VBM. Moreover, it can be seen that the least contribution to the total DOS towards VBM is by the Na-3s states. This feature of the Na atom is also observed in the inorganic double perovskites, Cs$_2$NaBX$_6$ (B=Bi, Sb; X= I, Br, Cl) investigated by (Shuai & Ma, 2016). The authors, observed that although the Na atom is important in the formation of the double perovskite crystal, the Na *s* orbital is not involved in the composition of VBM.

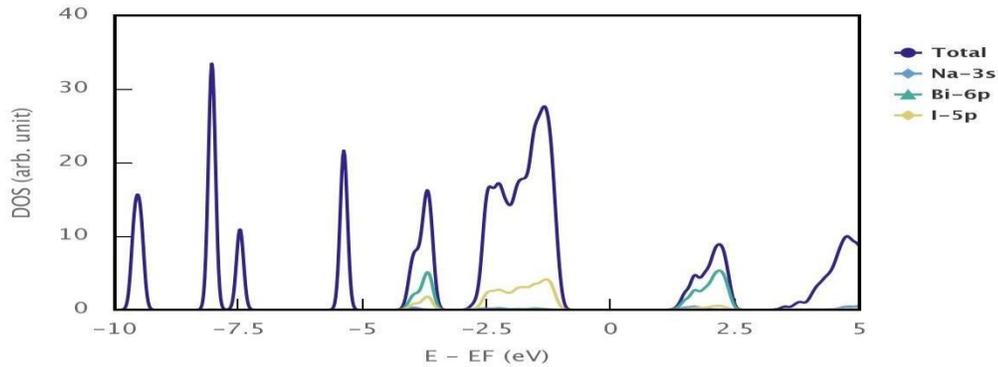

**Figure 4.** Calculated total and projected DOS of (CsMA)NaBiI$_6$ perovskite shows that the contribution to the total DOS towards VBM is by the Bi-6p states and I-5p states. The contribution to the total DOS towards CBM is by the I-5p states while the Na-3s states gave a partial contribution to the total DOS towards VBM.

### 3.3 Optical properties
The calculated dielectric functions, absorption coefficients, real parts of the refractive indices of (CsMA)NaBiX$_6$ (X = Cl, Br, I) are presented in Figures 5(a-c).

(a)

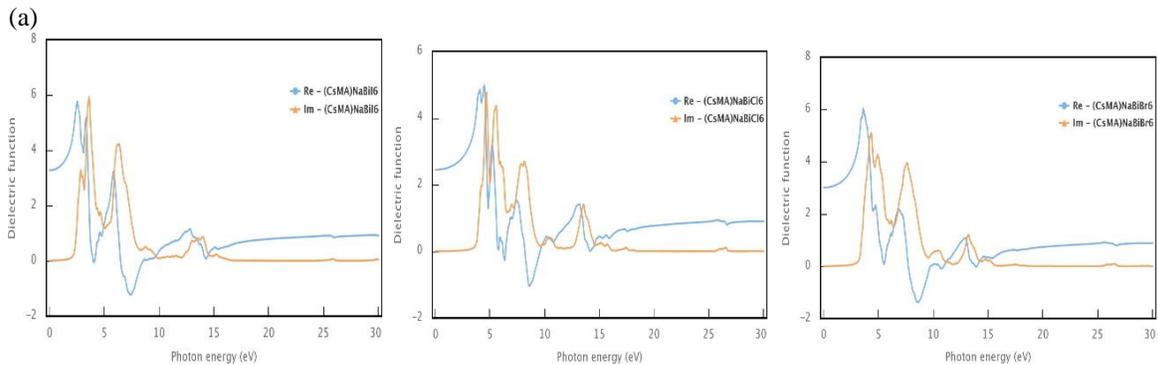

(b)

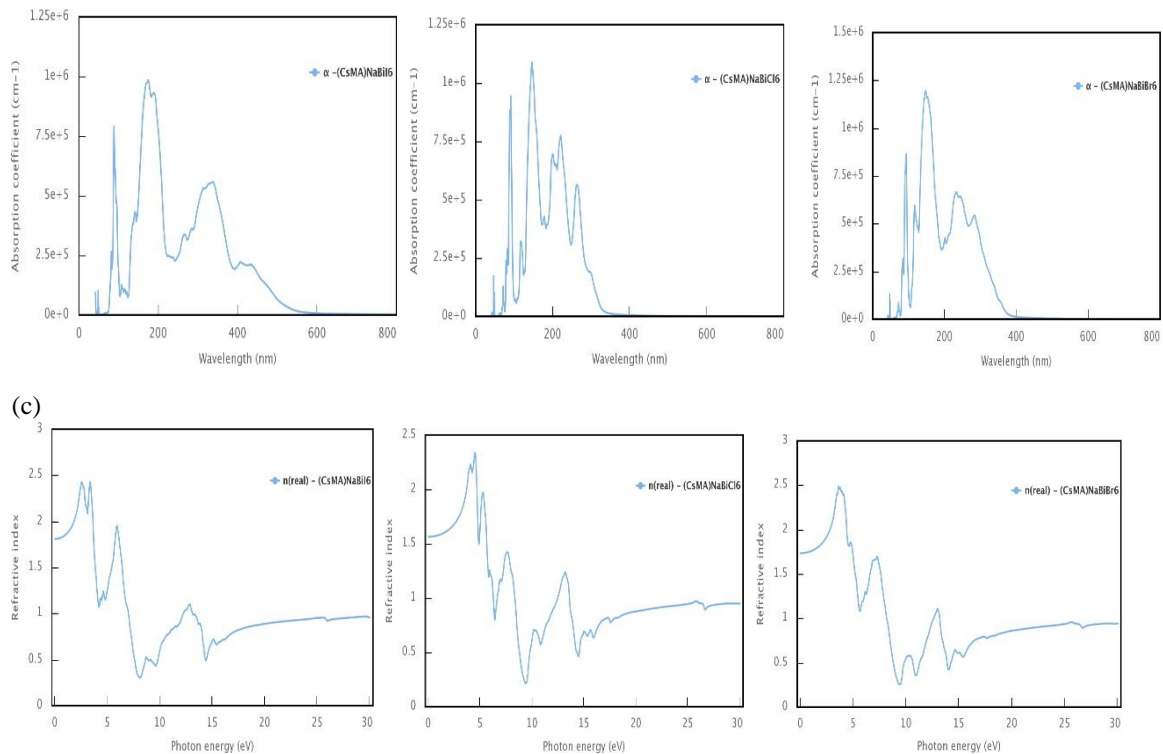

**Figure 5.** The optical properties of double perovskites (CsMA)NaBiX$_6$ (X = Cl, Br, I) along the incident electromagnetic radiation of energy from 0 to 30 eV. (**a**) Calculated dielectric function (**b**) Calculated absorption coefficient (**c**) Calculated refractive index.

The following optical properties of the hybrid double perovskites, namely, dielectric function, refractive index as well as the absorption coefficient were investigated. The dielectric function and refractive indices from Fig. 5 (a, c), were analysed for the photon energy between 0 and 30 eV; while the absorption coefficients were observed for the wavelength from 0 to 800 nm as in Fig. 5(b). From Fig. 5(a), the dielectric function intensity of the imaginary part of all samples varies in height from Iodine incorporated compound at 3.5 eV phonon energy to 3.8 eV and 4.5 eV for Bromine and Chlorine incorporated respectively. This trend is in conformity with the obtained Bandgap values for all the obtained samples; implying both the dielectric constant of the phonon energy and bandgap of these three analysed incorporated halogens follow same trend in value (3.5<3.8<4.5) eV for I$_6$<Br$_6$<Cl$_6$ respectively as observed by (Rajivgandh, Govindan Nadar Chackaravarthi et al., 2022).

The imaginary parts of the dielectric function of the considered semiconductor compounds were investigated due to the fact that the absorptive behaviour of a material depends on the imaginary part of the dielectric function (Shuai & Ma, 2016). The material (CsMA)NaBiI$_6$ depicts the highest value for the imaginary dielectric function of all the investigated compounds for solar radiation. This high value of the dielectric function of the (CsMA)NaBiI$_6$ material for solar radiation suggests that it is a potential optoelectronic material.

In general, the optoelectronic performance of a material depends on high dielectric constant or dielectric function at zero frequency. This is because materials with large dielectric constant can hold large amount of charge over a long period of time thus enhancing the optoelectronic performance (Johnson, A, Gbaorun, F and Ikyo, 2022; Ramanathan & Khalifeh, 2021). Among the investigated hybrid double perovskites in this study, the highest value of the dielectric constant of 3.24 obtained for (CsMA)NaBiI$_6$ material shows similarity with 4.43 obtained for the (CsMA)NaSbI$_6$ material (Johnson, A, Gbaorun, F and Ikyo, 2022). This implies that other optical properties of the hybrid double perovskite (CsMA)NaBiI$_6$ material could be a suitable substitute to Pb-based hybrid perovskites for optoelectronic applications. This is because the absorption coefficient of a material is a key determinant of its light harvesting capacity and plays a significant role in its application for optoelectronics.

Results of the present study, in all showed that the investigated materials have high absorption coefficient with the highest absorption obtained (in the order of $10^6 cm^{-1}$) with (CsMA)NaBiI$_6$. Moreover, as can be inferred from Figure 5b, the absorption coefficient decreases with increasing wavelength and is not uniform even if the photons have an energy greater than the band-gap energy as observed by (Jin, 2013). This is as a result of the fact that the probability of a photon being absorbed depends on the likelihood of interacting with an electron. Additionally, in Figure 5(c), the refractive index for (CsMA)NaBiI$_6$ showed the maximum peak (2.50) over a broad spectrum (2-4 eV) which implies that the (CsMA)NaBiI$_6$ material can be suitable in optoelectronic applications such as liquid crystal display (LCDs), organic-light emitting diode(OLED), quantum dot light emitting diode (QDLED) televisions which require materials withhigh refractive index (>1.50) (Garner, 2019; Johnson, A, Gbaorun, F and Ikyo, 2022).

### 3.4 Thermal properties

Table 2. Calculated formation energy, H(kJ/mol) of (CsMA)NaBiX$_6$ (X = Cl, Br, I)

| Compounds | H(kJ/mol) |
|---|---|
| (CsMA)NaBiI$_6$ | -4,706.0 |
| (CsMA)NaBiCl$_6$ | -5,126.4 |
| (CsMA)NaBiBr$_6$ | -4,955.3 |

As can be inferred from Table 2, the calculated formation energies of the considered hybrid double perovskites are presented. The formation energy of a material is the change of enthalpy when 1 mole of a compound is formed from its constituents' elements. According to (Roknuzzaman et al., 2019), a negative formation energy of a material reveals the stability of a material. This stability increases with increasing negativity of the formation energy. Lookingat the results in Table 2, the formation energies of the I-based compounds show least negativity compared with that of Br or Cl-based compounds, hence, the most stable. However, the resultsof the formation energies of the investigated compounds show that they are stable and can be utilized for different optoelectronic applications.

## 4 Conclusion

The investigation of the structural, electronic, optical properties and stability of 3 hybrid organic– inorganic double perovskites, (CsMA)NaBiX$_6$ (MA= methylammonium; X=I, Br, Cl) based on first-principle DFT/LDA method shows that the investigated compounds are semiconductors with tunable bandgaps characteristics that can be used in devices like light emitting diodes. Findings also reveal that the(CsMA)NaBiI$_6$ hybrid organic-inorganic double perovskite depicts superior optoelectronic properties such as high absorption coefficient, high refractive index compared with the other compounds in the study as shown by the LDA. In addition, the high absorption coefficient, high refractive indices and dielectric constants of the investigated materials posits that they are potential Pb-free materials for optoelectronic applications including photovoltaic cells.

**Acknowledgement**

The authors wish to acknowledge the Cloud Computing Interface of Materials Square and the Computational Facility of the Centre for Food Technology and Research (CEFTER), the World Bank Africa Centre of Excellence in Benue State University, Makurdi, Nigeria.

**Author Contributions**
All authors have contributed equally.

**Funding**
The authors received no funding for this work

**Data Availability**
The authors confirm the data of this study are available within the article.

# Declaration

**Conflict of interest**
The authors declare that they have no competing interests.

**Ethics approval**
Not applicable for that section.

**Informed consent**
All authors confirm their participation in this paper.

**Consent for publication**
All authors accept the publication rules applied by the journal.